\documentclass[11pt]{article}
\usepackage{amsfonts}
\usepackage{amsmath}

\newcommand{\ket}[1]{\mbox{$| #1 \rangle$}}
\newcommand{\braket}[2]{\mbox{$\langle #1 | #2 \rangle$}}

\def\squareforqed{\hbox{\rlap{$\sqcap$}$\sqcup$}}
\def\qed{\ifmmode\squareforqed\else{\unskip\nobreak\hfil
\penalty50\hskip1em\null\nobreak\hfil\squareforqed
\parfillskip=0pt\finalhyphendemerits=0\endgraf}\fi}
\newtheorem{theorem}{Theorem}
\newtheorem{lemma}[theorem]{Lemma}
\newenvironment{proof}{\begin{trivlist}\item[]{\flushleft\bf Proof }}
{\qed\end{trivlist}}

\DeclareMathSymbol{\itTheta}{\mathalpha}{letters}{"02}
\DeclareMathSymbol{\leqslant}{\mathrel}{AMSa}{"36}  

\newcommand{\subgroup}{\leqslant}          
\newcommand{\cs}[1]{{\langle #1 \rangle}}  

\title{Hidden Subgroup States are Almost Orthogonal}

\author{Mark Ettinger\,%
\thanks{\,\mbox{NIS--8}, \mbox{MS~B230}, 
Los Alamos National Laboratory, Los Alamos, NM~87545, USA.
Email: \texttt{\boldmath ettinger$\mathchar"40$lanl.gov}.}\\%
{\protect\small\sl LANL\/}%
\and
Peter H{\o}yer\,%
\thanks{\,BRICS, Department of Computer Science, 
University of Aarhus, \mbox{DK--8000} \mbox{{\AA}rhus~C}, Denmark.
Email: \texttt{\boldmath hoyer$\mathchar"40$brics.dk}.}\\%
{\protect\small\sl BRICS\/}\,%
\thanks{\,Basic Research in Computer Science, 
Centre of the Danish National Research Foundation.}
\and
Emanuel Knill\,%
\thanks{\,\mbox{CIC--3}, \mbox{MS~B265}, 
Los Alamos National Laboratory, Los Alamos, NM~87545, USA.
Email: \texttt{\boldmath knill$\mathchar"40$lanl.gov}.}\\%
{\protect\small\sl LANL\/}}

\date{{\normalsize January 14, 1999}}

\begin{document}

\maketitle

\begin{abstract}
It~is well known that quantum computers can efficiently find a hidden 
subgroup~$H$ of a finite Abelian group~$G$.  This implies that after only 
a polynomial (in~$\log |G|$) number of calls to the oracle function,
the states corresponding to different candidate subgroups have 
exponentially small inner product.  
We~show that this is true for noncommutative groups also.  
We~present a quantum algorithm which identifies a hidden subgroup of 
an arbitrary finite group~$G$ in only a linear (in~$\log |G|$) 
number of calls to the oracle function. 
This is exponentially better than the best classical algorithm. 
However our quantum algorithm requires an exponential amount of time, 
as in the classical case.
\end{abstract}


\section{Introduction}
A~function~$f$ on a finite group (with an arbitrary range)
is called {\em H-periodic\/} if
$f$ is constant on left cosets of~$H$.  
If~$f$ also takes distinct values on 
distinct cosets we say $f$ is {\em strictly} $H$-periodic.  Furthermore we
call $H$ the {\em hidden subgroup\/} of $f$.  Throughout we assume $f$ is
efficiently computable.  Utilizing the quantum Fourier
transform  a quantum computer can identify $H$ in time 
polynomial in~$\log |G|$.
The question has been repeatedly raised as to whether this may
accomplished for finite non-Abelian groups~\cite{EH99,Kitaev95,RB98}.

This time bound implies that only a polynomial (in~$\log |G|$) 
number of calls to the oracle function are necessary to identify~$H$.  
The main result of this paper
is that this more limited result is also true for non-Abelian groups.
In other words there exists a quantum algorithm which {\em
informational theoretically\/} determines the hidden subgroup
efficiently.  One may also view this as a quantum state
distinguishability problem and from this perspective one may say that
this quantum algorithm efficiently distinguishes among the given possible
states.  This result may be seen as a generalization of the
results presented in~\cite{EH99} although in that work the unitary transform
was also efficiently implementable.  The quantum algorithm presented
here requires exponential time.  An important open question is whether
this may be improved.  Even if a time efficient quantum algorithm does
exist one must also inquire as to the complexity of postprocessing the
resulting information.  For example in the case of the dihedral group
presented in~\cite{EH99}, although the hidden subgroup is information
theoretically determined in a polynomial number of calls to the oracle,
it is not known how to
efficiently postprocess the resulting information to identify~$H$.

\begin{theorem}[Main]\label{thm:main}
Let $G$ be a finite group, $f$ an oracle function on~$G$ which is
strictly $H$-periodic for some subgroup $H \subgroup G$.
Then there exists a quantum algorithm that calls the oracle function
$4 \log|G| +2$ times and outputs a subset $X \subseteq G$ 
such that $X=H$ with probability at least \mbox{$1 - 1/|G|$}.
\end{theorem}

\section{The Quantum Algorithm}
Let $G$ be a finite group, $H \subgroup G$ a subgroup, 
and $f$ a function on $G$ which is strictly $H$-periodic. 
For any subset $X = \{x_1,\dots,x_m\} \subseteq G$, let
$\ket{X}$ denote the normalized superposition
$\frac{1}{\sqrt{m}}\big(\ket{x_1} + \cdots + \ket{x_m}\big)$. 
The Hilbert Space~$\mathcal{H}$ in which
we work has dimension $|G|^m$ and has an orthonormal
basis indexed by the elements of the \mbox{$m$-fold} direct product
$\{\ket{(g_1,\dots,g_m)} \mid g_i \in G\}$.  
The first step in our quantum algorithm is to prepare the state 
\begin{equation}\label{eq:initial}
\frac{1}{\sqrt{|G|^m}} \sum_{g_1,\dots,g_m \in G}
{\ket{g_1,\dots,g_m} \,\ket{f(g_1),\dots,f(g_m)}}.  
\end{equation}
We show below that picking $m = 4 \log |G| +2$ 
allows us to identify~$H$ with exponentially small error probability.

By observing the second register we obtain a state~\ket{\Psi} which is a 
tensor product of random left cosets corresponding to the 
hidden subgroup~$H$.  Let
$\ket{\Psi} = \ket{a_{1}H} \otimes \ket{a_{2}H} \otimes \cdots \otimes
\ket{a_{m}H}$
where $\{a_1,\dots,a_m\} \subseteq G$.
Further, for any subgroup $K \subgroup G$ and 
any subset $\{b_1,\dots,b_m\} \subseteq G$, define 
\begin{equation}
\ket{\Psi(K,\{b_i\})} =
\ket{b_{1}K} \otimes \ket{b_{2}K} \otimes \cdots \otimes \ket{b_{m}K}.
\end{equation}  
The key lemma, stated formally below,
is that if $K \not\subgroup H$ then $\braket{\Psi}{\Psi(K,\{g_i\})}$ is
exponentially small.  

Let $\mathcal{H}_K$ be the subspace of $\mathcal{H}$ spanned by all
the vectors of the form $\ket{\Psi(K,\{b_i\})}$ for all subsets
$\{b_i\}.$  Let $P_K$ be the projection operator onto $\mathcal{H}_K$
and let $P_{K}^\perp$ be the projection operator onto the orthogonal
complement of $\mathcal{H}_K$ in $\mathcal{H}.$  Define the
observable $A_K = P_K - P_K^\perp$.  Choose an
ordering of the elements of $G$, say, $g_1,g_2,\dots,g_{|G|}.$

The algorithm mentioned in Theorem~\ref{thm:main} works as follows.
We~first apply $A_\cs{g_1}$ to $\ket{\Psi}$, 
where $\cs{g} \subgroup G$ denotes 
the cyclic subgroup generated by \mbox{$g \in G$}.
If~the outcome is~$-1$ then we know that $g_1 \not\in H$, 
and if the outcome is~$+1$ then we know that
$g_1 \in H$ with high probability.  
We~then apply $A_\cs{g_2}$ to the 
state resulting from the first measurement.  
Continuing in this manner we test
each element of $G$ for membership in $H$ by sequentially applying
$A_\cs{g_2}$, $A_\cs{g_3}$ and so on to the resulting states 
of the previous measurements.
(Of~course if we discover $g \in H$ then we know that, say,
$g^2 \in H$ and we can omit the test $A_\cs{g^2}$.)  
We~prove below that each measurement alters the state
insignificantly with high probability, 
implying that by the application of the final operator $A_\cs{g_{|G|}}$ 
we have, with high probability, identified exactly which 
elements of~$G$ are in~$H$ and which are not.

\begin{lemma}\label{lm:key}
Let $K \subgroup G$.  If $K \not\subgroup H$ 
then $\langle \Psi|P_K |\Psi \rangle \leq \frac{1}{2^m}$.
If~$K \subgroup H$ then $\langle \Psi|P_K |\Psi \rangle = 1$.
\end{lemma}

\begin{proof}
Let $|H \cap K| = d$.  Notice that for all
$g_1,g_2 \in G$ we have either $|g_1H \cap g_2K| = d$ or $|g_1H \cap
g_2K| = 0$.  This implies that if $|g_1H \cap g_2K| = d$ then
$\braket{g_1H}{g_2K} = d/\sqrt{|H||K|}$.  
Therefore
\begin{equation*}
\braket{\Psi}{\Psi(K,\{b_i\})} = 
\begin{cases}\Big(\frac{d}{\sqrt{|H||K|}}\Big)^m 
  & \mbox{ if $|H \cap K| = d$}\\ 0 & \mbox{ if $|H \cap K| = 0$.}
\end{cases}
\end{equation*}
There exist exactly $(|H|/d)^m$ vectors of the form
$\ket{\Psi(K,\{b_i\})}$ such that $\braket{\Psi}{\Psi(K,\{b_i\})}$
is nonzero.  Hence, $\langle \Psi |P_K | \Psi \rangle =
\big(\frac{|H|}{d}\big)^m \big(\frac{d^2}{|H||K|}\big)^m 
= \big(\frac{d}{|K|}\big)^m$.
If $K \not\subgroup H$ then $d/|K| \leq 1/2$, and if
$K \subgroup H$ then \mbox{$d = |K|$}.
\end{proof} 

Let $\ket{\Psi_0} = \ket{\Psi}$.
For $1 \leq i \leq |G|$, define the unnormalized states
\begin{equation*}
\ket{\Psi_i} = 
\begin{cases}
\,P_\cs{g_i} \;\ket{\Psi_{i-1}} & \mbox{ if $g_i \in H$}\\
\,P_\cs{g_i}^\perp \;\ket{\Psi_{i-1}} & \mbox{ if $g_i \not\in H$.}
\end{cases}
\end{equation*}
Then $\braket{\Psi_i}{\Psi_i}$ equals the probability that
the algorithm given above answers correctly whether $g_j \in H$ for 
all $1 \leq j \leq i$.  
Now, for all $0 \leq i \leq |G|$, 
let $\ket{E_i} = \ket{\Psi} - \ket{\Psi_i}$.

\begin{lemma}\label{lm:error}
For all $0 \leq i \leq |G|$, we have 
$\braket{E_i}{E_i} \leq \frac{i^2}{2^m}$.
\end{lemma}

\begin{proof}
We~prove this by induction on~$i$.
Since $\ket{\Psi_0} = \ket{\Psi}$ by definition, $\ket{E_0} = 0$.
Now, suppose that $\braket{E_i}{E_i} \leq \frac{i^2}{2^m}$.
On~the one hand, if $g_{i+1} \in H$, then 
$\ket{\Psi_{i+1}} = P_{\cs{g_{i+1}}} \big(\ket{\Psi} - \ket{E_{i}}\big)
= \ket{\Psi} -  P_{\cs{g_{i+1}}} \ket{E_{i}}$.
Hence $\braket{E_{i+1}}{E_{i+1}} \leq \braket{E_i}{E_i} \leq 
\frac{i^2}{2^m}$.
On~the other hand, if $g_{i+1} \not\in H$, then
$\ket{\Psi_{i+1}} = P_{\cs{g_{i+1}}}^\perp \big(\ket{\Psi} - \ket{E_{i}}\big)
= \ket{\Psi} - P_{\cs{g_{i+1}}} \ket{\Psi} - P_{\cs{g_{i+1}}}^\perp 
\ket{E_{i}}$.
By~Lemma~\ref{lm:key}, we then have 
$\braket{E_{i+1}}{E_{i+1}}^{1/2} \leq 
\frac{1}{2^{m/2}} + \braket{E_i}{E_i}^{1/2} \leq \frac{i+1}{2^{m/2}}$.
\end{proof}

Since $\ket{\Psi_{|G|}} = \ket{\Psi} - \ket{E_{|G|}}$ 
and $\braket{E_{|G|}}{E_{|G|}} \leq \frac{|G|^2}{2^m}$
by the above lemma, we obtain the following lower bound
for correctly determining all the elements of~$H$.

\begin{lemma}
$\braket{\Psi_{|G|}}{\Psi_{|G|}} \,\geq  1 - \frac{2 |G|}{2^{m/2}}$.
\end{lemma}

By choosing $m = 4 \log|G| +2$, the main theorem follows directly.

\section{Conclusion}
We have shown that there exists a quantum algorithm that discovers a
hidden subgroup of an arbitrary finite group in $O(\log|G|)$ calls to
the oracle function.  This is possible due to the geometric fact that
the possible pure states corresponding to different possible subgroups
are almost orthogonal, i.e. they have exponentially small inner
product.  Equivalently stated, there exists a measurement, a POVM, that
distinguishes among the possible states in a Hilbert Space of
dimension $|G|^m$ where $m = O(\log|G|)$.  The open question remains in
regard to the
existence of a POVM which not only distinguishes among the states but
is efficiently implementable and the resulting information is
efficiently postprocessable.

\end{document}